# Electronic Structures of Atomic Silicon Dimer Wires as a Function of Length

Furkan M. Altincicek[1], Lucian Livadaru[1], Christopher C. Leon[1*], Taras Chutora[1], Max Yuan[1], Roshan Achal[2], Jeremiah Croshaw[1], Jason Pitters[3], Robert Wolkow[1,2]

[1]*Department of Physics, University of Alberta, Edmonton, Alberta, T6G 2E1, Canada*

[2]*Quantum Silicon Inc., Edmonton, Alberta, T6G 2M9, Canada*

[3]*National Research Council of Canada, Edmonton, Alberta, T6G 2M9, Canada*

[*]*Current address: Département de chimie, Université Laval, Québec, Québec G1V 0A6, Canada*

Correspondence to: rwolkow@ualberta.ca

## Abstract

Bare silicon dimers on hydrogen-terminated Si(100) have two dangling bonds. These are atomically localized regions of high state density near to and within the bulk silicon band gap. We studied bare silicon dimers as monomeric units. Silicon dimer wires are much more stable than wires composed of individual dangling bonds. Dimer wires composed of 1 to 5 dimers were intentionally fabricated and characterised by STM techniques combined with density functional theory to provide detailed insights into geometric and electronic structure. Structural and dynamic qualities displayed by short wires were shown to be similar to the characteristics of a relatively long 37 dimer wire. Rather than adding two states into the band gap, experiment and theory reveal that each dimer adds one empty state into the gap and one filled state into the valence bands. Coupling among these states provides a conduction pathway with small bulk coupling.

## Introduction

Atomic-scale structures have never been more accessible than they are today thanks to advances in scanning tunneling microscopy (STM) and its variants. Its unique ability to resolve and manipulate atoms has made it an invaluable tool for this purpose [1, 2]. In the case of H–Si(100)–2×1, this semiconducting surface has a dimerized repeating pattern and one monolayer of hydrogen that acts as an atomically thin resist well-suited for these types of studies [3–9]. The surface is used as a canvas for atomic scale patterning on which the H–Si bond allows for the selective removal of individual hydrogen atoms using the STM tip. This procedure leaves behind silicon dangling bonds (DBs) [3] which are atomic orbitals that can be used to create functional structures [10–18]. More complicated molecular-like orbitals result from combining atomic orbitals in close proximity. A pair of DBs situated on the same silicon dimer is termed a "bare dimer" which is the basic structural unit considered in this study. Henceforth, all references to "dimer" refer to this structure (as opposed to the H-capped dimer). They are key to constructing complex atomic-scale structures such as wires [10, 16], artificial molecules [5, 13], and binding sites for dopant precursors [17, 18]. An extensive review has recently been published that encompasses various examples of such DB patterns on this surface [19].

To gain a comprehensive understanding of the electronic properties of bare silicon dimers, we conducted orbital analyses on the 1 to 3 dimer wire segments. Fabrication was done by intentionally creating one, two, and three units of dimers through STM tip manipulation on the H-Si(100) surface of degenerately doped p-type silicon at 4.5 K. This high level of doping is helpful for avoiding dopant freeze-out at 4.5 K, which hinders scanning. Our findings show that for every dimer, there is one filled state in the valence band and one empty state in the band gap. In particular, these states hybridize and cause the apparent band gap to close from the conduction band side. Due to tip-induced band bending, these states can appear convolved with the conduction band during STM measurements, and tunneling occurs through electron-hole recombination. Modeling through density function theory (DFT) was consistent with STM measurements.

## Methods

*Equipment*

All measurements were taken with a commercially available Omicron low-temperature STM (LT-STM). Experiments were conducted at 4.5 K in a chamber with a $5\times10^{-11}$ Torr base pressure. The silicon sample was cleaved from a degenerately boron-doped (0.005 $\Omega\cdot$cm) wafer. It was degassed at 600 °C overnight and was flash annealed to 1250 °C several times with a direct current heating. After the last flash the temperature of the sample was dropped to 330 °C to terminate the surface with a hydrogen monolayer. During this stage, molecular hydrogen ($H_2$) was leaked into the preparation chamber and cracked to atomic hydrogen (H) with a tungsten filament heated to ~1800 °C [3] for 2 minutes while maintaining $1\times10^{-6}$ Torr pressure (120 Langmuirs). The STM tip was cut from W(111) wire and electrochemically etched in 2M NaOH solution [20]. It was later sharpened to a single atom apex using a home-built Field Ion Microscopy (FIM), where helium and nitrogen gasses were used to image and etch the tip, respectively [21]. During STM measurements, controlled contacts were performed on DB patches in order to alter or improve the tip, whenever necessary.

*Measurements*

Constant height measurements were performed with a height setpoint of 1.8 V and 20 pA on H-Si atoms with a homogeneous background. For lock-in measurements, 50 mV bias modulation amplitude and 585 Hz frequency were used. All images are presented as forward scans, except for Figure 4q, for which forward and backward scans were averaged. Constant height dI/dV line spectroscopies were performed with a grid spacing and energy resolution of 50 pm and 20 mV respectively. Line spectroscopies were taken separately for each dimer wire and stitched together in Figure 5.

*Theory*

Density Functional Theory (DFT) calculations were performed with the AMS2021 [22, 23] program using the GGA-PBE [24] exchange-correlation (XC) functionals. All dimer wires were subject to geometry optimizations using a nano-crystal model, $Si_{228}H_{178}$, H-capped on all sides. A double-zeta basis set was used for all the atoms, except for a region consisting of the clean dimer atoms and the 4 Si atoms directly bonded to it, for which a double-zeta polarized basis set was used. This latter region was allowed full flexibility during geometry optimization, while the other atoms were fixed. In order to better approximate the real system, the optimized wire structures were then translated into a periodic slab model, with a (3×8) surface supercell consisting of $Si_{576}H_{142}$ and with 12 Si layers in depth, top and bottom being H-capped. Single-point slab calculations were then performed for each optimized dimer wire case. For a judicious calculation, the supercell was divided in two regions, according to the importance for the present study. The first "central" region consists of the atoms of the dimer row where the dimer wire is placed, excepting the Si atoms adjacent to the border of the supercell. The remaining atoms were allocated to the "support" region. A double-zeta polarized basis set was used for all the atoms (both regions), but we made use of a frozen core for the "support" region, while the "central" region was given an all-electron treatment. Gamma-point only was used for k-space.

## Results and Discussions

Bare silicon dimers were fabricated either by removing H atoms individually or simultaneously in pairs with bias pulses [9, 13]. Pulsing was done from a set point of 1.6 V and 50 pA, to a bias of 2.4 V for a duration of 50 ms with no tip feedback. Such bias pulses cause vibrational excitations through inelastic tunneling which ruptures the hydrogen-silicon bond [25–27]. The smallest three silicon dimer wire segments are shown in Figure 1a. We define them to be of length 1, 2 and 3, which are equivalent to compact DBs arrays of size 2×1, 2×2 and 2×3. While our focus here is on structure evolution among the smallest dimer wires, Figure 1c shows a 37 dimer long wire (2×37) to demonstrate that given a defect-free area, these wires can be extended to any desired length.

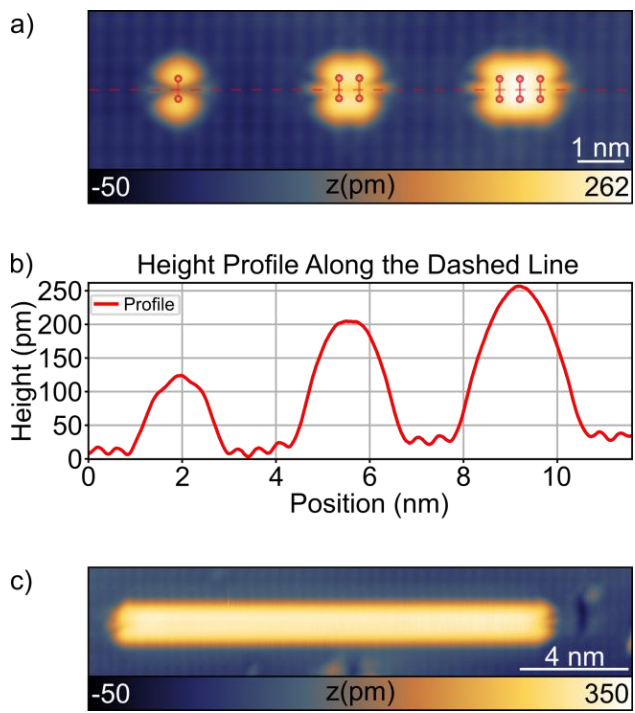

**Figure 1) a)** A STM image of silicon dimers on the H–Si(100)–2×1 surface (1.8 V, 50 pA constant current). Atomic positions of the dimers are indicated with superimposed ball and stick models. **b)** The height profile across the line is indicated with a dashed red line in Figure 1a. **c)** A 37-dimer long atomic wire.

The image shows a clear contrast between the crystalline array of atomically resolved H atoms in the background in dark blue and the very pronounced dangling bonds situated on the dimers that protrude from the surface in bright yellow. This morphology reflects the relatively low electron density on the H-terminated area and the high electron density on the dimer structures at the probed energy.

These 3 dimer wires coexisted on the surface and were imaged simultaneously to ensure a common tip structure was used to image all three units. We focus on this reduced set of structures because they contain features that are common to all dimer wires, *i.e.,* all dimer wires contain dimers that are either in a "middle" or at a terminal position. The first is the fundamental and smallest unit of all dimer wires. The second is a wire segment with two distinct ends. The third contains an additional middle dimer. Studying these structures in this sequence revealed that longer dimer structures exhibit higher protrusions compared to the shorter fragments. This is clearly seen in Figure 1a and 1b, in which the brightness of the dimer wire segments increases with intensity as a function of length. Adding additional dimers to the single bare dimer, increases tunneling pathways and causes the 3 dimer wire to appear brighter at the center dimer [16]. The image of H-terminated areas near this structure causes those to show up brighter compared to hydrogen atoms near a single dimer also, as shown in the height profile in Figure 1b. This is a pinning effect. When the tip is positioned over the H-terminated region the nearby Si dimers resist the upward band bending field effect of the tip by taking on a reduced negative charge. This effectively creates a larger bias between tip and sample, resulting in more current.

To better understand the effects of grouping dimer units together, dI/dV measurements, which are proportional to local density of states (LDOS), were measured using a lock-in amplifier. In each case, spectroscopies were taken in the middle of each dimer unit. For the 2 and 3 dimers wires, the multiple spectroscopies were averaged into the single curves which are shown in Figure 2a.

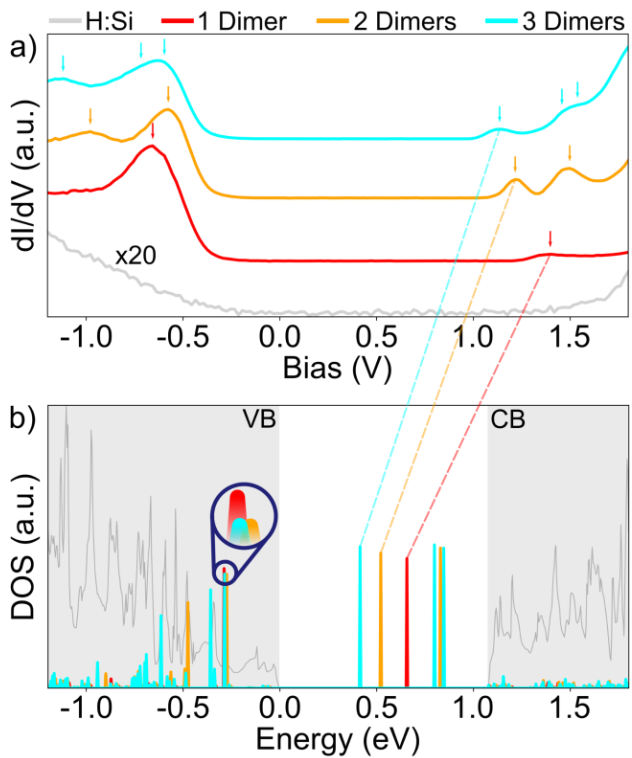

**Figure 2) a)** Experimental dI/dV spectroscopies and **b)** theoretical partial DOS plots of 1, 2 and 3 dimers, showing the addition of empty and filled states with each added dimer. The curves in (a) were color coded to match the corresponding states in (b). Wire states in the bandgap are visually matched with their experimental counterparts with dashed lines, with the energy mismatch due to TIBB. The gray background curve in (b) indicates the total DOS of the pristine model slab in absence of the wires.

These experimental LDOS for 1, 2 and 3 dimers, as well as for H–Si, are given in Figure 2a. The measurements revealed the existence of broad features at energies below -0.5 V and above +1.0 V, at the positions indicated with arrows. These features are assigned to filled (occupied) and empty (unoccupied) electronic states, respectively. Despite the filled states being resonant with the valence band states, they are resolvable as a result of being positioned on the surface and thereby overlapping more strongly with the tip than bulk states [28]. Although the experimental dI/dV measurements were unable to fully resolve some individual states, a comparison with calculated DFT projected density of states in Figure 2b reveals that they do originate from distinct molecular-like orbitals.

This comparison requires some care because tip-induced band bending (TIBB) shifts the apparent position of states [29]. TIBB is especially pronounced for states furthest away from the sample Fermi level [29-31]. DFT estimates the unperturbed energy levels in the absence of TIBB. This accounts for the discrepancy in energetic positions of the empty state features indicated in Figure 2a and Figure 2b which are connected by dashed lines between the two subpanels. The filled states are exposed to smaller TIBB in comparison to empty states because they are much closer to the p-type sample Fermi level.

Of particular interest is the case for 3 dimers shown in light blue in Figure 2a. In the one and two dimer long wires, each dimer adds a negative bias and a positive bias peak to the dI/dV spectroscopies, however this is not observed for the 3 dimer long wire. The DFT calculations reveal the existence of close spaced HOMO/HOMO-1 and LUMO-1/LUMO-2 orbitals that are unresolvable in experiment. The full set of molecular orbitals marked with arrows in Figure 2a are considered further in the discussion of Figure 4.

The DFT results additionally reveal that the filled and empty states are energetically positioned in the valence band and band gap respectively. These states can be further identified as a sequence of HOMO-n and LUMO+n states with n = 0, 1, 2. It is significant that the filled states overlap with the valence band and thus, do not cause any shrinkage of the apparent band gap. Instead, it is the energetic positions of the empty states that effectively narrows the apparent band gap specifically from the conduction band side.

The tunneling mechanisms responsible for producing the features seen in Figure 2 can be rationalized with a set of heuristic energy level diagrams as shown in Figure 3 where the case of three dimers is shown. Figures 3a-b show the tunneling mechanism for the dimer wire filled states that are aligned with the Si valence band, and Figures 3c-d show the mechanism for the empty dimer wire states located in the Si band gap. The sample Fermi level, indicated with a thick dashed line, is drawn slightly below the valence band edge to indicate a degenerately doped p-type sample [32]. The alignment of the filled states with the valence band allows resonant tunneling of electrons from the wire to the tip as shown in Figures 3a-b. Measuring the empty states involves a different mechanism due to their energy in the band gap which prohibits resonant tunneling. In this case (Figures 3c-d), electrons are injected into the wire from the tip, to then inelastically tunnel to the valence band where they combine with abundant holes [30].

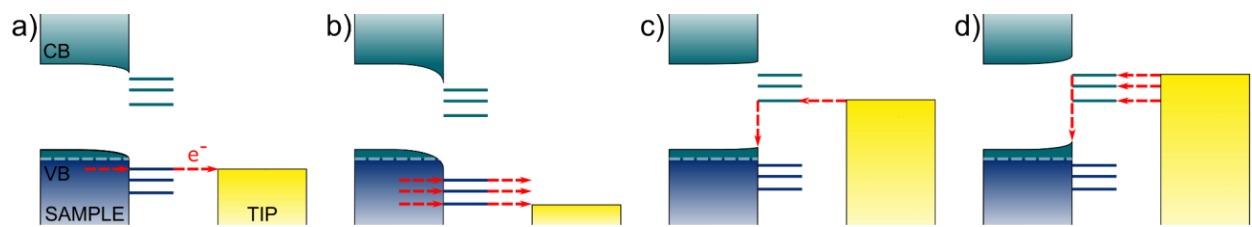

**Figure 3)** Tunneling mechanism for imaging the **a)** HOMO **b)** HOMO-2 **c)** LUMO and **d)** LUMO+2 for three dimers. Navy and teal indicate filled and empty states, respectively.

Additionally, the positions of the filled and empty states with respect to the Fermi level have an impact on the measured energy of the states. The filled states are accessed at relatively low tip voltages, resulting in minor amounts of TIBB. This is visible in Figure 2, where the experimental and theoretical filled states have similar energies. On the other hand, the much higher tip voltages needed to access the empty states, causes significantly more TIBB and creates a larger energy mismatch between the experimental and theoretical empty states in the same Figure indicated with dashed lines.

At each energetic position of the features identified in Figure 2, we recorded a spatial density of states map with STM (Figures 4a-f and 4m-r) and compared these measurements with DFT calculations (Figures 4g-l and 4s-x). Of interest are the regions of high electron density because they may correlate with chemical reactivity depending on the reactant and might be used as intentionally made binding sites [17, 18]. Figure 4 shows that filled states are most pronounced along the centres of the dimer rows whereas empty states are split above and below this center line. These features lengthen with each additional dimer and generally exhibit regularities in their spatial distribution. In the empty states, addition of dimers adds one more state in the empty states that resembles particle-in-a-box-like states (HOMO-n), while for the filled states this is not observed due to the aforementioned split with respect to the center line.

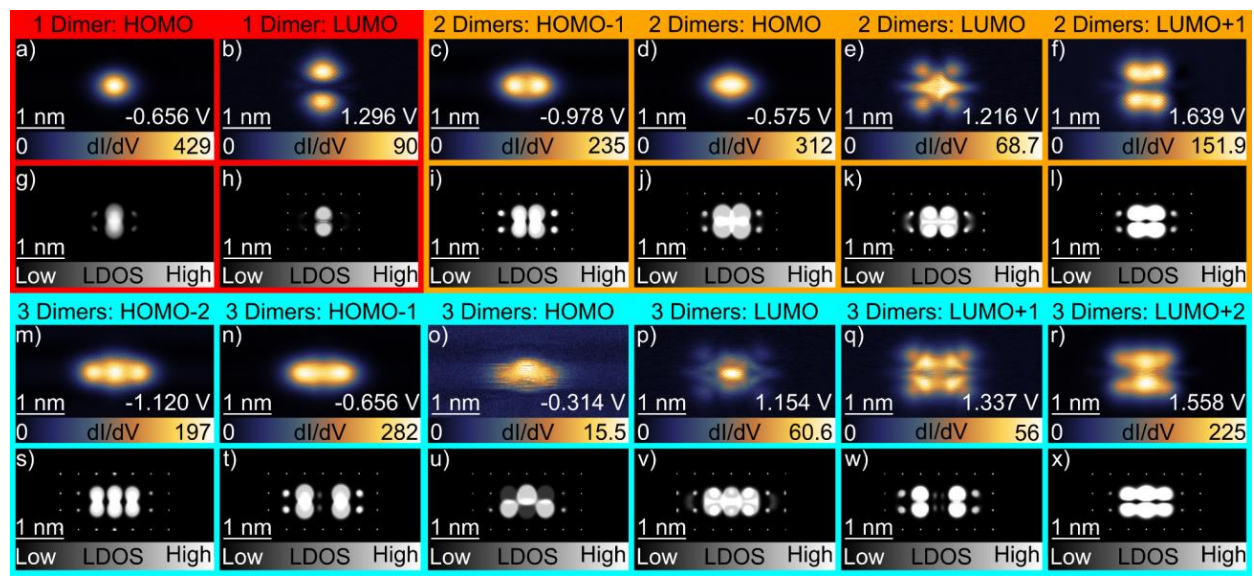

**Figure 4)** Experimental **(a-f and m-r)** and simulated **(g-l and s-x)** images of filled and empty molecular orbitals of one, two and three dimers. Corresponding experimental and theoretical images were put one under the other for ease of comparison and the titles of the molecular orbitals are indicated on the images. Frame colours are matched with Figure 2. A blue-yellow colour scale (in arbitrary units) is used to show STM images (Figure 4a-f, m-r) and a black-white scale is used for simulated DFT images (Figure 4g-l, s-x). Image processing details can be found in Methods. Tip height is set on a hydrogen atom at 1.8 V, 20 pA.

Figure 4 shows electron density of the structures that is time averaged as dimers rapidly switch among two buckled states. To approximate this effect, DFT results were presented as equal weighted averages of two buckled configurations, except for Figure 4u, where the averaging was weighted 80-20%. This is because a close inspection of Figure 4o (3 dimers, HOMO) reveals a streaky and reduced symmetry buckled texture. The buckling is characteristic of the Si(100)–2×1 reconstruction in which adjacent dimers are tilted oppositely [32-34]. In the case of Figure 4o, the effect of inelastic electron tunneling is just large enough to cause switching of buckled dimers. The 1 and 2 dimer structures are believed to be flipping too rapidly to detect with our limited (~1 kHz) bandwidth [16].

Switching rates for buckled configurations decrease as more dimers are added, due to increased strain-mediated coupling. At room temperature it has been shown that dimers fluctuate freely, unless an asymmetrically placed surface defect pins adjacent dimers into only one buckled configuration [32]. It is evident that STM current effectively causes local heating of the surface through inelastic processes.

We observe no evidence of a symmetric dimer configuration in the experiment or theory. A symmetric dimer has only a fleeting existence as a transition state among stable degenerate buckled configurations. The dominant configurations during transport are the two degenerate ground states (which are perfectly alternating buckled geometries, and are the mirror reflection of each other w.r.t. the surface-perpendicular median plane along the length of the wire) and the lowest lying excited states exhibiting 1 or more kinks – defects separating two or more segments with perfect zig-zag geometry. Switching between the two degenerate ground states is intermediated by the formation of excited states having 1 kink that easily shifts among different locations along the wire. All of the above configurations exhibit electronic states that are delocalized along the wire, facilitating transport. Therefore, conduction is by a mixture of coherent (when the wire remains in the ground state) and incoherent transport (when the wire is excited in one of the low-lying states).

A previous study [10] demonstrated that extending dimer wires reduces the apparent band gap of the structures, causing longer dimer wires to show more metallic character on n-type at 77 K. That work revealed numerous key features that inspired our further exploration of the subject. Moreover, as atomic device manufacturing is as likely to require p-type substrates as n-type, we investigated dimers on p-type silicon at 4.5 K. Using the same dI/dV spectroscopic imaging approaches along lines at the centers of dimers as given in Figure 5. The most striking difference between the previous study and our current work is that, in the latter, we lack metallic high-conductivity in the dopant-depleted subsurface region. Despite the expected increase in the band gap towards lower temperatures, the main cause of this difference is the position of the sample Fermi level. It is apparent that our heating process during sample preparation resulted in fewer dopants in the near-surface region, resulting in larger tip-induced band bending and a larger apparent gap.

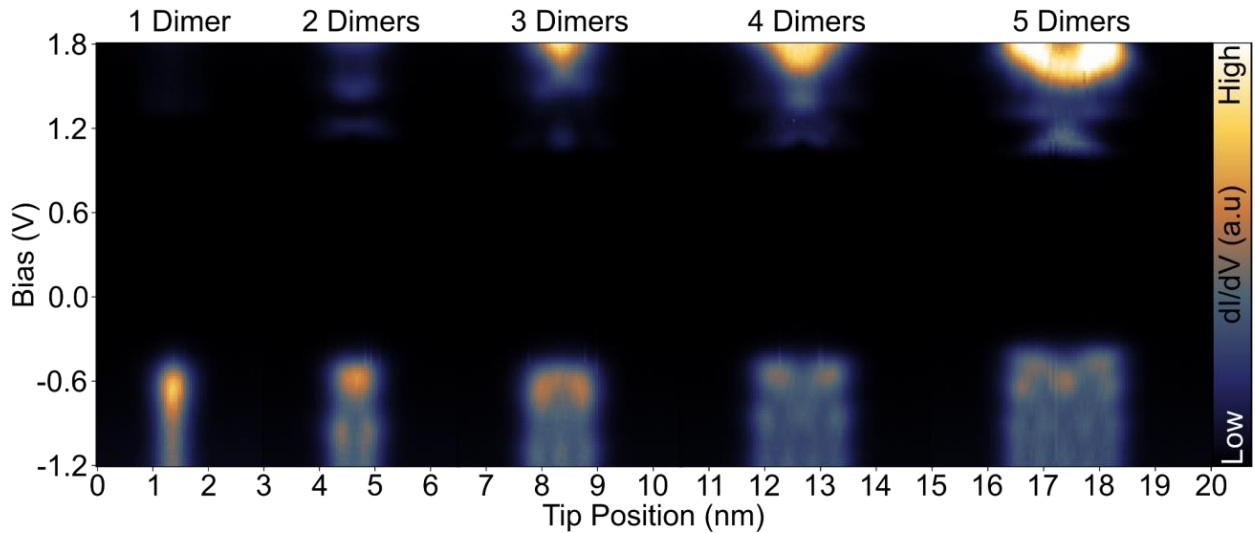

**Figure 5)** dI/dV line spectroscopies on various lengths of dimer wires. dI/dV signal was acquired along the centerline of 1, 2, 3, 4 and 5 dimers at a height set on a H atom at 1.8 V 20 pA.

Figure 5 also shows the standing wave pattern in the filled states, confirming that addition of every dimer does create an equal number of filled states. Similar trends for empty states are difficult to observe in this data set since the point spectroscopies were taken along the centerline of dimers where some empty states do not show high electron density as seen in Figures 4b,f,r.

## Conclusions

In conclusion, we studied silicon dimer wires on the H–Si(100)–2×1 surface with STM and DFT. With each additional dimer, one filled state is introduced to the valence band and one empty state enters the band gap. This reduces the apparent band gap from the conduction band side as the length of the structure increases. These features exhibit regularities in their density of states in terms of energetic position and spatial distribution. It is our hope that these findings may aid in the progress of silicon-based nanotechnology. Specifically, unoccupied dimer wire states which are isolated from the bulk states may serve to conduct electrons between atomically defined points. Both occupied and unoccupied surface states might be used to direct attachment of adsorbates. These observations highlight the importance of understanding the electronic properties of bare silicon dimers for the development of atomic-scale devices.

## Acknowledgements

Funding has been supplied by the National Research Council of Canada, Alberta Innovates Technology Futures, Natural Sciences and Engineering Research Council of Canada, and the Digital Research Alliance of Canada.

# References

1) Binnig, G.; Rohrer, H.; Gerber, Ch.; Weibel, E. Surface Studies by Scanning Tunneling Microscopy. Physical Review Letters 1982, 49, 57–61. https://doi.org/10.1103/PhysRevLett.49.57

2) Eigler, D. M.; Schweizer, E. K. Positioning Single Atoms with a Scanning Tunnelling Microscope. Nature 1990, 344, 524–526. https://doi.org/10.1038/344524a0

3) Lyding, J. W.; Shen, T.-C.; Abeln, G. C.; Wang, C.; Tucker, J. R. Nanoscale Patterning and Selective Chemistry of Silicon Surfaces by Ultrahigh-Vacuum Scanning Tunneling Microscopy. Nanotechnology 1996, 7, 128–133. https://doi.org/10.1088/0957-4484/7/2/006

4) McEllistrem, M.; Allgeier, M.; Boland, J. J. Dangling Bond Dynamics on the Silicon (100)-2×1 Surface: Dissociation, Diffusion, and Recombination. Science 1998, 279, 545–548. https://doi.org/10.1126/science.279.5350.545

5) Schofield, S. R.; Studer, P.; Hirjibehedin, C. F.; Curson, N. J.; Aeppli, G.; Bowler, D. R. Quantum Engineering at the Silicon Surface Using Dangling Bonds. Nature Communications 2013, 4. https://doi.org/10.1038/ncomms2679

6) Wolkow, R. A.; Livadaru, L.; Pitters, J.; Taucer, M.; Piva, P.; Salomons, M.; Cloutier, M.; Martins, B. V. Silicon Atomic Quantum Dots Enable beyond-CMOS Electronics. Field-Coupled Nanocomputing 2014, 33–58. https://doi.org/10.1007/ 978-3-662-43722-3_3

7) Taucer, M.; Livadaru, L.; Piva, P. G.; Achal, R.; Labidi, H.; Pitters, J. L.; Wolkow, R. A. Single-Electron Dynamics of an Atomic Silicon Quantum Dot on the H-Si(100)-(2x1) Surface. Physical Review Letters 2014, 112. https://doi.org/10.1103/physrevlett.112.256801

8) Huff, T. R.; Labidi, H.; Rashidi, M.; Koleini, M.; Achal, R.; Salomons, M. H.; Wolkow, R. A. Atomic White-out: Enabling Atomic Circuitry through Mechanically Induced Bonding of Single Hydrogen Atoms to a Silicon Surface. ACS Nano 2017, 11, 8636–8642. https://doi.org/10.1021/acsnano.7b04238

9) Achal, R.; Rashidi, M.; Croshaw, J.; Churchill, D.; Taucer, M.; Huff, T.; Cloutier, M.; Pitters, J.; Wolkow, R. A. Lithography for Robust and Editable Atomic-Scale Silicon Devices and Memories. Nature Communications 2018, 9. https://doi.org/10.1038/ s41467-018-05171-y

10) Naydenov, B.; Boland, J. J. Engineering the Electronic Structure of Surface Dangling Bond Nanowires of Different Size and Dimensionality. Nanotechnology 2013, 24, 275202. https://doi.org/10.1088/0957-4484/24/27/275202

11) Kawai, H.; Neucheva, O.; Yap, T. L.; Joachim, C.; Saeys, M. Electronic Characterization of a Single Dangling Bond on N- and P-Type Si(001)-(2 × 1):H. Surface Science 2016, 645, 88–92. https://doi.org/10.1016/j.susc.2015.11.001

12) Engelund, M.; Godlewski, S.; Kolmer, M.; Zuzak, R.; Such, B.; Frederiksen, T.; Szymonski, M.; Sánchez-Portal, D. The Butterfly – a Well-Defined Constant-Current Topography Pattern on Si(001):H and Ge(001):H Resulting from Current-Induced Defect Fluctuations. Physical Chemistry Chemical Physics 2016, 18, 19309–19317. https://doi.org/10.1039/c6cp04031d

13) Wyrick, J.; Wang, X.; Namboodiri, P.; Schmucker, S. W.; Kashid, R. V.; Silver, R. M. Atom-by-Atom Construction of a Cyclic Artificial Molecule in Silicon. Nano Letters 2018, 18, 7502–7508. https://doi.org/10.1021/acs.nanolett.8b02919

14) Huff, T.; Labidi, H.; Rashidi, M.; Livadaru, L.; Dienel, T.; Achal, R.; Vine, W.; Pitters, J.; Wolkow, R. A. Binary Atomic Silicon Logic. Nature Electronics 2018, 1, 636–643. https://doi.org/10.1038/s41928-018-0180-3

15) Achal, R.; Rashidi, M.; Croshaw, J.; Huff, T. R.; Wolkow, R. A. Detecting and Directing Single Molecule Binding Events on H-Si(100) with Application to Ultradense Data Storage. ACS Nano 2020, 14 (3), 2947–2955. https://doi.org/10.1021/acsnano.9b07637

16) Croshaw, J.; Huff, T.; Rashidi, M.; Wood, J.; Lloyd, E.; Pitters, J.; Wolkow, R. A. Ionic Charge Distributions in Silicon Atomic Surface Wires. Nanoscale 2021, 13, 3237–3245. https://doi.org/10.1039/d0nr08295c

17) Wyrick, J.; Wang, X.; Namboodiri, P.; Kashid, R. V.; Fei, F.; Fox, J.; Silver, R. Enhanced Atomic Precision Fabrication by Adsorption of Phosphine into Engineered Dangling Bonds on H–Si Using STM and DFT. ACS Nano 2022, 16, 19114–19123. https://doi.org/10.1021/acsnano.2c08162

18) Campbell, Q.; Ivie, J. A.; Bussmann, E.; Schmucker, S. W.; Baczewski, A. D.; Misra, S. A Model for Atomic Precision P-Type Doping with Diborane on Si(100)-2×1. The Journal of Physical Chemistry C 2021, 125, 481–488. https://doi.org/10.1021/acs.jpcc.0c08919

19) Pitters, J.; Croshaw, J.; Achal, R.; Livadaru, L.; Ng, S.; Lupoiu, R.; Chutora, T.; Huff, T.; Walus, K.; Wolkow, R. A. Atomically precise manufacturing of silicon electronics. ACS Nano 2024, 18 (9), 6766–6816. https://doi.org/10.1021/acsnano.3c10412

20) Kazinczi, R.; Szõcs, E.; Kálmán, E.; Nagy, P. Novel Methods for Preparing EC STM Tips. Applied Physics A: Materials Science & Processing 1998, 66. https://doi.org/10.1007/s003390051197

21) Rezeq, M.; Pitters, J.; Wolkow, R. Tungsten Nanotip Fabrication by Spatially Controlled Field-Assisted Reaction with Nitrogen. The Journal of Chemical Physics 2006, 124. https://doi.org/10.1063/1.2198536

22) Velde, G. T.; Bickelhaupt, F. M.; Baerends, E. J.; Guerra, C. F.; Van Gisbergen, S. J. A.; Snijders, J. G.; Ziegler, T. Chemistry with ADF. Journal of Computational Chemistry 2001, 22 (9), 931–967. https://doi.org/10.1002/jcc.1056

23) ADF 2019, SCM, Theoretical Chemistry, Vrije Universiteit, Amsterdam, The Netherlands, http://www.scm.com

24) Perdew, J. P.; Burke, K.; Ernzerhof, M. Generalized gradient approximation made simple. Physical Review Letters 1996, 77 (18), 3865–3868. https://doi.org/10.1103/physrevlett.77.3865

25) Shen, T.-C.; Wang, C.; Abeln, G. C.; Tucker, J. R.; Lyding, J. W.; Avouris, P.; Walkup, R. E. Atomic-Scale Desorption through Electronic and Vibrational Excitation Mechanisms. Science 1995, 268, 1590–1592. https://doi.org/10.1126/science.268.5217.1590

26) Soukiassian, L.; Mayne, A. J.; Carbone, M.; Dujardin, G. Atomic-Scale Desorption of H Atoms from the Si(100)-2×1:H Surface: Inelastic Electron Interactions. Physical Review B 2003, 68. https://doi.org/10.1103/PhysRevB.68.035303

27) Randall, J. N.; Lyding, J. W.; Schmucker, S.; Von Ehr, J. R.; Ballard, J.; Saini, R.; Xu, H.; Ding, Y. Atomic Precision Lithography on Si. Journal of Vacuum Science & Technology B: Microelectronics and Nanometer Structures Processing, Measurement, and Phenomena 2009, 27, 2764–2768. https://doi.org/10.1116/1.3237096

28) Rashidi, M.; Taucer, M.; Ozfidan, I.; Lloyd, E.; Koleini, M.; Labidi, H.; Pitters, J. L.; Maciejko, J.; Wolkow, R. A. Time-Resolved imaging of negative differential resistance on the atomic scale. Physical Review Letters 2016, 117 (27). https://doi.org/10.1103/physrevlett.117.276805

29) Feenstra, R. M.; Dong, Y.; Semtsiv, M. P.; Masselink, W. T. Influence of Tip-Induced Band Bending on Tunnelling Spectra of Semiconductor Surfaces. Nanotechnology 2006, 18, 044015. http://doi.org/10.1088/0957-4484/18/4/044015

30) Kittel, C. Introduction to Solid State Physics; Wiley, 2004. ISBN: 978-0-471-41526-8.

31) Voigtländer, B. Scanning Probe Microscopy; Springer: Berlin, 2015. ISBN: 978-3-662-45239-4. https://doi.org/10.1007/978-3-662-45240-0

32) Wolkow, R. A. Direct Observation of an Increase in Buckled Dimers on Si(001) at Low Temperature. Physical Review Letters 1992, 68, 2636–2639. https://doi.org/10.1103/PhysRevLett.68.2636

33) Hitosugi, T.; Heike, S.; Onogi, T.; Hashizume, T.; Watanabe, S.; Li, Z.-Q.; Ohno, K.; Kawazoe, Y.; Hasegawa, T.; Kitazawa, K. Jahn-Teller Distortion in Dangling-Bond Linear Chains Fabricated on a Hydrogen-Terminated Si(100)-2×1 Surface. Physical Review Letters 1999, 82, 4034–4037. https://doi.org/10.1103/PhysRevLett.82.4034

34) Ren, X.-Y.; Kim, H.-J.; Niu, C.-Y.; Jia, Y.; Cho, J.-H. Origin of Symmetric Dimer Images of Si(001) Observed by Low-Temperature Scanning Tunneling Microscopy. Scientific Reports 2016, 6. https://doi.org/10.1038/srep27868